\documentclass{WileyChemistry-template}
\usepackage{chemformula}
\usepackage[T1]{fontenc} 
\usepackage{graphicx}
\usepackage{amsthm,amsmath}
\usepackage{tabularx}

\title{\raggedright Effect of Protonation and Deprotonation on Electron Transfer Mediated Decay and Interatomic Coulombic Decay}

\author{
\begin{minipage}{\textwidth}
	Ravi Kumar,\textsuperscript{[a]} Dr. Nayana Vaval,*\textsuperscript{[a]}
\end{minipage}
}

\newcommand{\affiliation}{
\begin{itemize}

\item[{[a]}] R. Kumar, Dr. N. Vaval*\\
Academy of Scientific and Innovative Research, CSIR-Human Resource Development Center (CSIR-HRDC) Campus, Postal staff College Area, Ghaziabad, Uttar Pradesh, 201002, India \\Electronic Structure Theory Group, Physical and Materials Chemistry Division, CSIR-National Chemical Laboratory, Pune, 411008, India\\
E-mail: np.vaval@ncl.res.in

\end{itemize}
}

\renewcommand{\abstract}{Electronically excited atoms or molecules in an environment are often subject to interatomic/intermolecular Coulombic decay (ICD) and/or electron transfer mediated decay (ETMD) mechanisms. A few of the numerous variables that can impact these non-radiative decay mechanisms include bond distance, the number of nearby atoms or molecules, and the polarisation effect. In this paper, we have studied the effect of protonation and deprotonation on the ionization potential (IP), double ionization potential (DIP), and lifetime (or decay width) of the temporary bound state in these non-radiative decay processes. We have chosen LiH-NH$_3$ and LiH-H$_2$O as test systems. The equation of motion coupled cluster singles and doubles method augmented by complex absorbing potential (CAP-EOM-CCSD) has been used in calculating the energetic position of the decaying state and the system's decay rate. Deprotonation of LiH-NH$_3$/LiH-H$_2$O either from the metal center (LiH) or from ammonia/water lowers the IP and DIP compared to the neutral systems.  In contrast,  protonation increases these quantities compared to neutral systems. The protonation closes the inner valence state relaxation channels for ICD/ETMD. For example, the decay of the O-2s/N-2s state stops in protonated systems (LiH$_2^+$-H$_2$O, LiH$_2^+$-NH$_3$, and LiH-NH$_4^+$). Our study also shows that the efficiency, i.e., the rate of ICD/ETMD, can be altered by protonation and deprotonation. It is expected to have implications for chemical and biological systems.
}

\begin{document}
\twocolumn[\vspace{-1.5cm}\maketitle\vspace{-1cm}
	\textit{\dedication}\vspace{0.4cm}]
\small{\begin{shaded}
		\noindent\abstract
	\end{shaded}
}

\begin{figure} [!b]
\begin{minipage}[t]{\columnwidth}{\rule{\columnwidth}{1pt}\footnotesize{\textsf{\affiliation}}}\end{minipage}
\end{figure}

\section*{Introduction}
\label{introduction}
\begin{figure}
  \includegraphics[width=\linewidth]{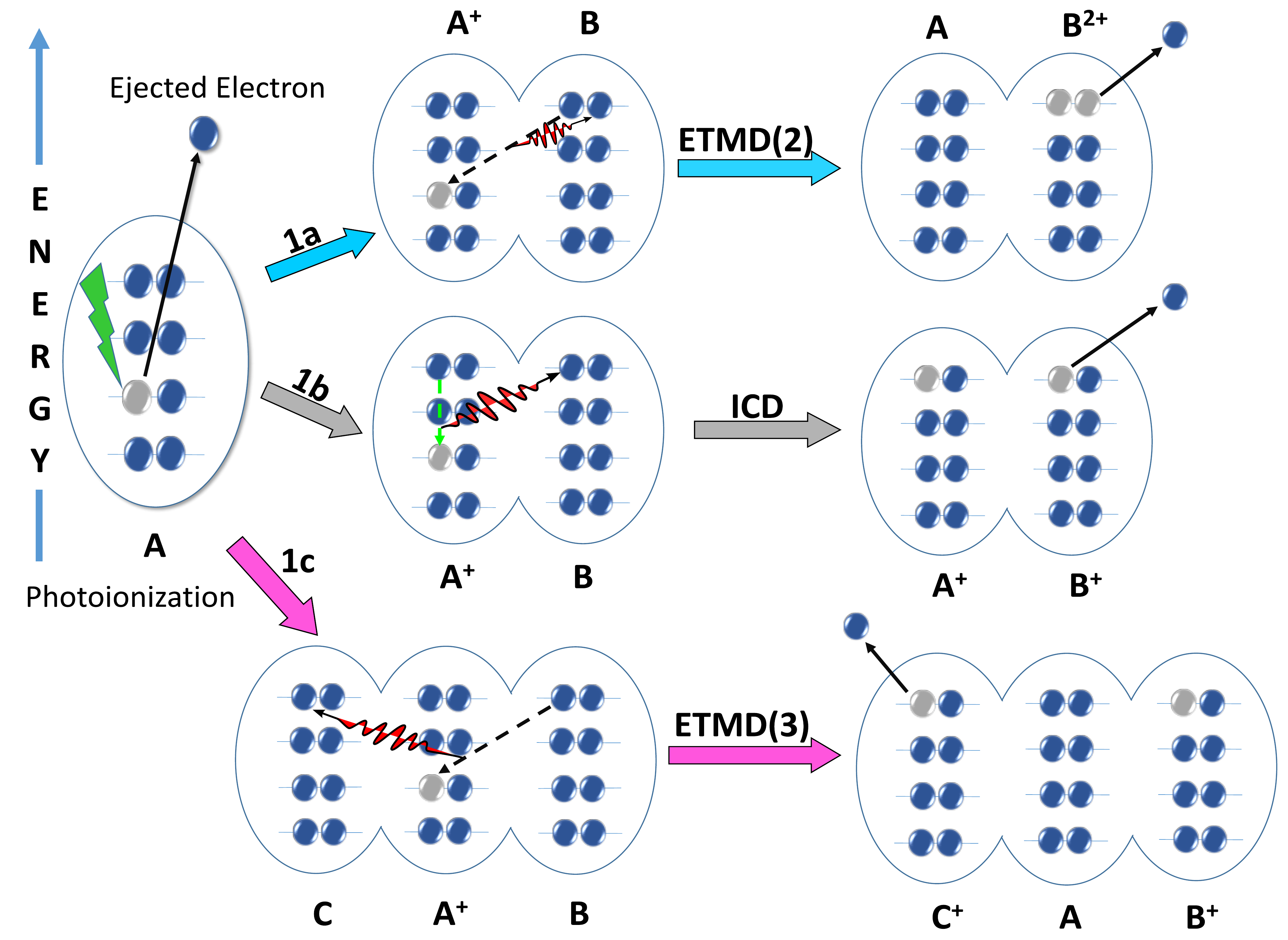}
  \caption{Visual representation of electronic decay processes. 
(1a.) In ETMD(2), an inner valence vacancy is filled by an outer valence electron of the neighboring molecule B, and another valence electron is emitted from the same molecule B leaving a final AB$^{2+}$ molecular state. 
(1b.) ICD, a valence electron from molecule A, fills the vacancy, and the released energy is transferred to an outer valence electron of neighboring molecule B. Thus, forming A$^+$B$^+$ as the final state.
(1c.) ETMD(3) process in which the vacancy in molecule A is filled by an outer valence electron of the neighboring molecule B, and ejection of the second electron occurs from another adjacent molecule C, leaving  AB$^+$C$^+$ as the final state. The number 2 and 3 represents the number of molecule involved in the ETMD process.}
  \label{icd-etmd}
\end{figure}

The knowledge on the importance of non-radiative processes like interatomic and intermolecular coulombic decay (ICD) \cite{ICD,ICD1}
and electron transfer mediated decay (ETMD) \cite{etmd,etmdhbond,etmdhbond1}
has substantially increased over the past two decades. ICD, first proposed by Cederbaum \textit{et al.}, \cite{ICD,ICD1} 
is a nonlocal and efficient decay mechanism that occurs at the femtosecond timescale. During this process, when an
ionized/excited system relaxes non-radiatively in an environment, an outer valence electron fills the vacancy in the inner valence,
resulting in a virtual photon emission which knocks out a valence electron from an adjacent atom or molecule. This situation leads to a 
Coulomb explosion due to the proximity of two positively charged species. The energy transfer happens quickly, i.e.
in a few to a hundred femtoseconds. ICD was initially studied theoretically
 \cite{ICDtheo,icdtheo1,icdtheo2} and experimentally \cite{ICDexpt,ICDexpt1,ICDexpt2,aq-etmd,etmd-ion,liq-water}
in hydrogen-bonded, \cite{icdhbond} and weakly bound rare gas clusters. \cite{icdrare,icdhene}
In ETMD, an electron is transferred from a neighboring atom to an initially ionized atom or molecule. 
The excess energy can knock out a second electron from the same donor atom/molecule in ETMD(2) process. 
While in ETMD(3), secondary electron ejection occurred from the second adjacent atom/molecule.
Since ETMD involves the transfer of electrons from a neighboring species to an initially ionized species, it is usually slower than the ICD. 
See figure \ref{icd-etmd}  for a better understanding of the difference between ICD and ETMD. The low energy electrons generated during ICD 
and ETMD have different kinetic energies. One can use this information to identify the specific local environment. 
Thus, the study of ICD and ETMD is essential to understand the decay of the ionized inner valence state in an environment.
Recent studies \cite{ravijctc,review} show that various factors influence the ICD/ETMD process, 
like the number of surrounding atoms or molecules, bond distance, geometry, and the medium in which 
energy/electron transfer happens. The environment can change the lifetime of a temporary bound state 
(TBS) undergoing ICD / ETMD by stabilizing or destabilizing it. Thus, environmental effects/properties
play a particularly relevant role. In this paper, we explore an important environmental effect:
the protonation and deprotonation effect on ICD and ETMD. Specifically, we will study the impact of 
protonation and deprotonation on ionization potential (IP), Double ionization potential (DIP), 
and decay width (or the lifetime of TBS). Recently, the effect of protonation and deprotonation
on IP and DIP in pure ammonia clusters \cite{lenzph} has been studied . Still, the impact of protonation and 
deprotonation on the lifetime of TBS has not been studied. We are the first ones to learn it through this study.

X <return>
Our aim is to understand the effect of protonation/deprotonation on the ICD/ETMD using alkali metal ions microsolvated in water and ammonia.
Understanding this may help us control the relevant process in a chemical environment. This tells us how a change in surrounding and protonation/deprotonation affects the decay process and lifetime of the TBS. The first step towards this will be doing these calculations accurately for a metal ion-water system and its protonated/ deprotonated form with the coupled-cluster methods, which scale as $N^{6}$. To get the decay width, we need to solve the CC equation thousands of times, which makes it computationally expensive. Thus, we wanted to study the smallest possible metal-related system. Therefore, we chose Lithium, although we know that the alkali metals Na$^{+}$, K$^{+}$, etc., are more relevant biologically. Also, as the first step towards this, gas-phase calculations are done. We may have to implement some approximate method for the actual interesting systems. We have discussed our finding and their possible reasons in the results and discussion section, followed by the Conclusion. Last, we have discussed the Computational and Theoretical details of the Equation of motion coupled cluster methods.

\section*{Results and Discussion}
  \label{results_discussion}

We will study the effect of protonation and deprotonation on the IP,  DIP, and lifetime of O-2s, N-2s, and Li-1s TBS in the gaseous state. We have chosen LiH-H$_{2}$O and LiH-NH$_3$ as test systems for our study. Protonation and deprotonation can occur either from LiH (metal center) or H$_2$O/NH$_3$. Thus,
LiH$^{+}_2$-NH$_3$ and LiH-NH$_4^{+}$ will be formed by the protonation of LiH-NH$_3$ and Li$^{-}$-NH$_3$ and LiH-NH$^{-}_2$ by deprotonation.  
Similarly, for the LiH-H$_{2}$O system, we have Li$^-$-H$_{2}$O and LiH-OH$^-$
as deprotonated and LiH$_2^+$-H$_{2}$O only as a protonated system. Due to the instability of LiH-H$_3$O$^+$, global minima could not be found.
Understanding the impact of deprotonation and protonation on ICD and ETMD in microsolvated systems is a step toward helping us control these relevant and abundant processes in various chemical environments and may also be in biological systems. We will first study, analyze and then compare the effect of the protonation and deprotonation on IP  and DIP for LiH-NH$_3$, LiH-H$_2$O, along with their protonated and deprotonated systems. Later, we will examine the factors that affect TBS lifetime and whether it is possible to find a general trend in the lifetime of TBS on protonation and deprotonation.

\begin{figure}
  \includegraphics[width=\linewidth]{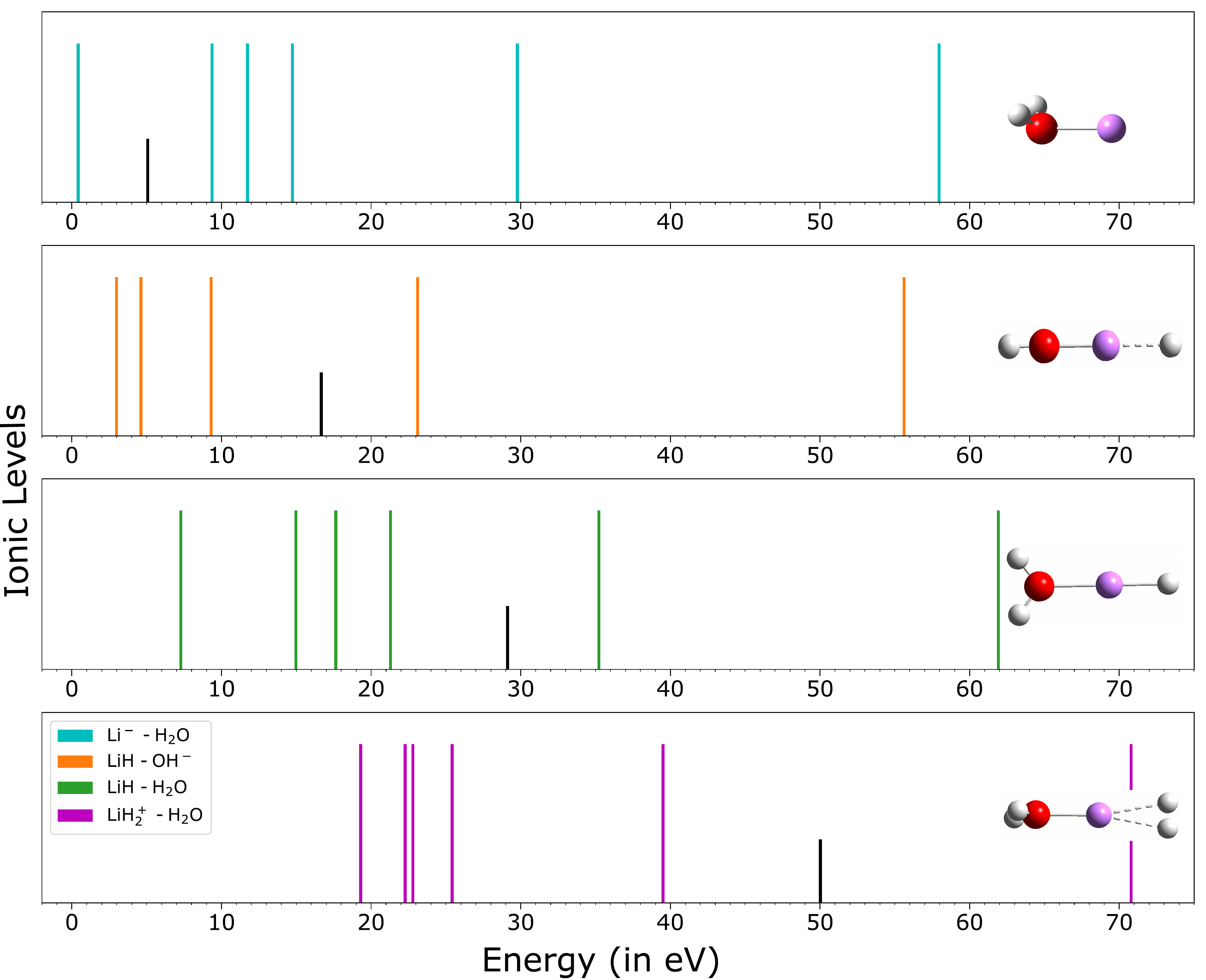}
  \caption{The IPs of neutral LiH-H$_2$O and its deprotonated and protonated species. The smaller black line denotes the lowest DIP value. The structure of the respective system is shown on the right of each panel. Atom color code : Red: Oxygen, Pink: Lithium, and White: Hydrogen. }
  \label{h2o-ip}
\end{figure}

\begin{figure}
  \includegraphics[width=\linewidth]{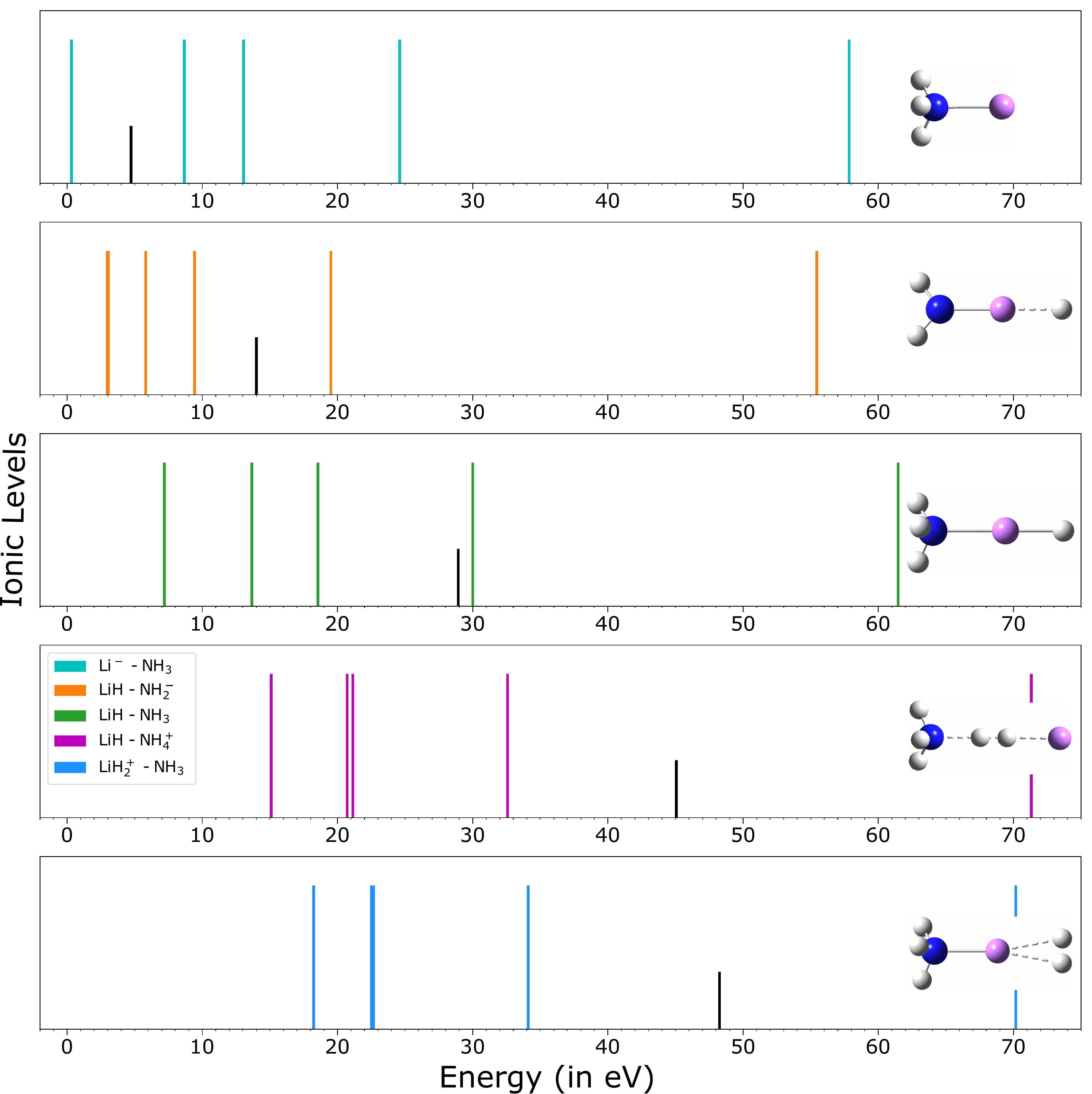}
  \caption{The IPs of deprotonated, protonated, and neutral LiH-NH$_3$. The smaller black line denotes the lowest DIP value. The structure of the respective system is shown on the right of each panel. Atom color code : Blue: Nitrogen, Pink: Lithium, and White: Hydrogen. }
  \label{nh3-ip}
\end{figure}

\subsection*{Effect of protonation and deprotonation on the IP and DIP}

Figures \ref{h2o-ip} and \ref{nh3-ip} display the all IP (except O-1s) and the lowest DIP values of the neutral, protonated, and deprotonated LiH-H$_2$O and LiH-NH$_3$ systems, respectively. 
We have written down a few observations from both figures (\ref{h2o-ip}nd and \ref{nh3-ip}rd) and will provide possible explanations for them. First, we observe the decrease in system's IP and DIP values with deprotonation and an increase after protonation. 
This is due to the decrease and increase in effective nuclear charge on deprotonation and protonation, respectively.
Protonation increases the effective nuclear charge (ENC)  while deprotonation decreases it. Electron ejection becomes challenging as ENC rises. As a result, we saw that IP and DIP value increased on protonation and decreased after protonation.
Second, from figure \ref{nh3-ip}, if we compare the IP values of the two deprotonated systems of LiH-NH$_3$, i.e., Li$^-$-NH$_3$ and LiH-NH$^-_2$.
The IP values for all the states of Li$^-$-NH$_3$ are higher than the corresponding LiH-NH$^-_2$, except for the IP of HOMO. HOMO's IP of LiH-NH$^-_2$ (3.06 eV) is more than HOMO's IP of Li$^-$-NH$_3$ (0.33 eV), which is just opposite of the rest of the IP trend. A similar trend is observed between deprotonated systems of LiH-H$_2$O.

 The IP values of HOMO  in the metal-centered deprotonated systems (Li$^-$-NH$_3$ and  Li$^-$-H$_2$O) almost drop to zero. The lowest IP (HOMO's IP) values of metal-centered deprotonated systems are so small that both systems may lose their outer-valance electron vibrationally. 
To know that, we have calculated these system's ionization energy (IE) and zero point energy. The IE value was found to be 0.54 eV and 0.46 eV for the Li$^{-}$-H$_{2}$O and Li$^-$-NH$_3$, respectively. But the zero point energy is 0.62 eV for Li$^{-}$-H$_{2}$O and 0.99 eV for Li$^-$-NH$_3$. It is clear that the zero-point energy of the metal-centered deprotonated systems is higher than its ionization energy. Thus, we conclude that the metal-centered deprotonated systems can lose electrons vibrationally in their ground state and are not electronically stable. Therefore, we will not further study the metal-centered deprotonated systems for the decay width. 
We have used the CCSD(T) method (a gold standard in theory) for the ionization energy and zero-point energy with the aug-cc-pVTZ basis set.\cite{ccpvtz} 

Till now, it’s clear that electrons can lose vibrationally, but it does not tell why the HOMO’s IP values of metal-centered deprotonated systems are so low than HOMO’s IP of their respective neutral systems and other deprotonated systems (LiH-NH$_2^-$ and LiH-OH$^-$). To understand that, we will discuss both systems' geometric and electronic structures before returning to the ionization spectra. Our primary focus will be the HOMO since HOMO does not fit the rest of the IP trend.

\begin{figure}
  \includegraphics[width=\linewidth]{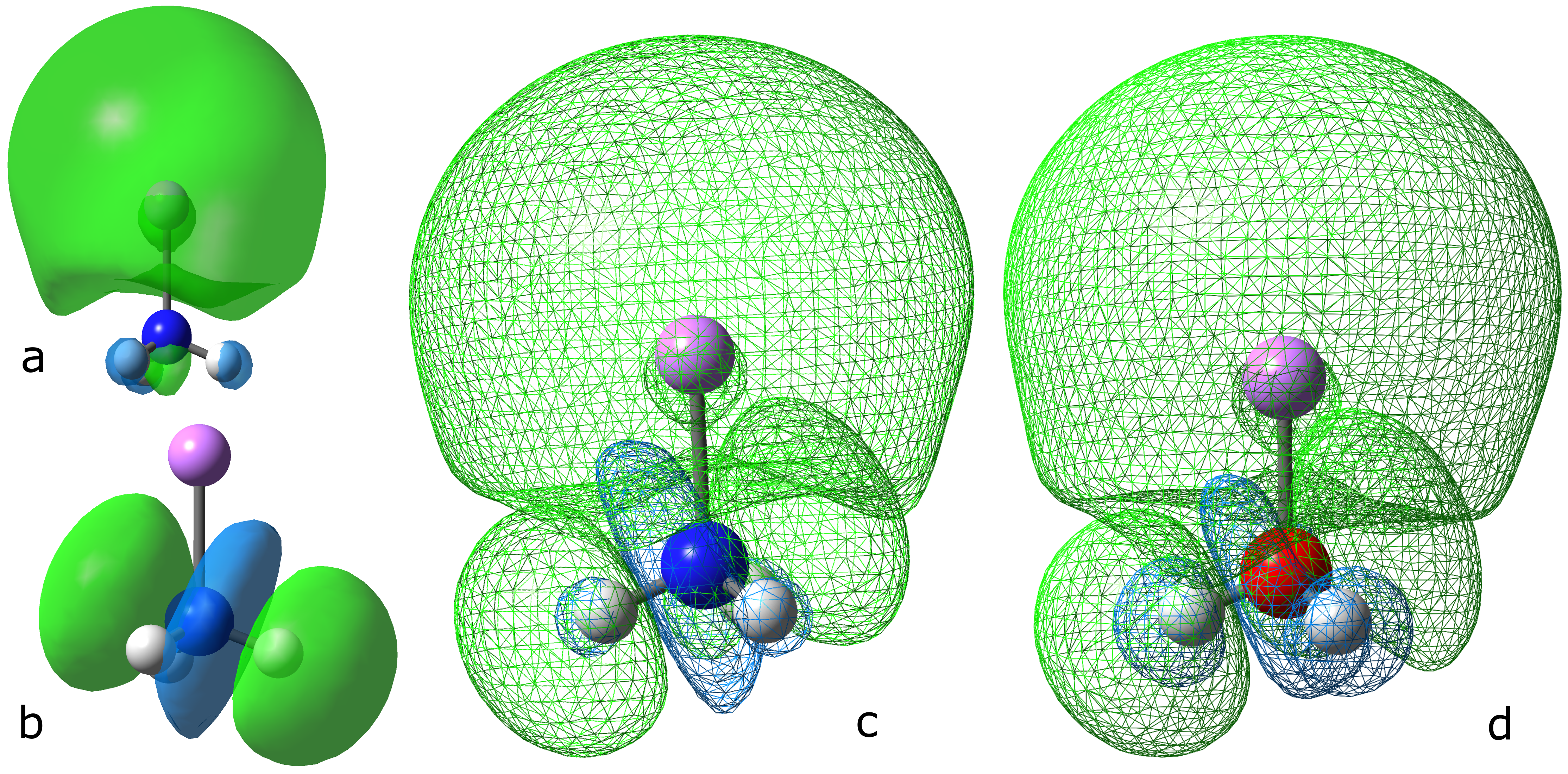}
  \caption{a and b show HOMO and LUMO in Li$^-$-NH$_3$. c shows HOMO and LUMO are together in one picture in Li$^-$-NH$_3$. d shows the same for Li$^-$-H$_2$O. This figure (c and d) demonstrates these orbitals partially overlap in space. Images are generated for 0.02 e\AA$^{-3}$ isosurface value. Atom color code-: Blue: Nitrogen, Red: Oxygen, Pink: Lithium, and White: Hydrogen.}
  \label{homo-lumo}
\end{figure}

We must know about the HOMO and its constituting atomic orbitals in neutral and metal-centered deprotonated systems to understand the changes clearly. Lithium-2s and hydrogen-1s atomic orbitals form the HOMO in neutral systems (LiH-NH$_3$ and LiH-H$_{2}$O). After deprotonation from the metal center, HOMO is mainly Li-2s in metal-centered deprotonated systems (Li$^-$-NH$_3$ and Li$^{-}$-H$_{2}$O). However, the HOMO of other deprotonated systems (LiH-NH$_2^-$/LiH-OH$^-$) is similar to the corresponding neutral systems. The LiH bond is ionic, so electron density is mainly towards the more electronegative atom in the Li-H bond, which is the hydrogen atom. Figure \ref{homo-lumo}(a,b, and c) shows the location of HOMO (Li-2s), the lowest unoccupied molecular orbital (LUMO) (N-H antibonding), and both in one frame for Li$^{-}$-NH$_{3}$. There is a partial overlapping in space between the HOMO and LUMO, which indicates the possibility of electron density transfer from HOMO to LUMO. 
Natural bond order (NBO) analysis proves this possibility, showing that the system stabilizes by 8.66$\times$3 = 25.98 kcal/mol by delocalizing the electron density to 3 degenerate N-H antibonding orbitals from HOMO. Since the HOMO of Li-2s is spherical, all three delocalizations can happen simultaneously. One more interaction stabilizing the system by 3.61 kcal/mol occurs between HOMO and Rydberg states (3s orbital) of N. Resulting in HOMO will have negligible electron density. Hence, the IP of HOMO is lowered dramatically to 0.33 eV in  Li$^{-}$-NH$_{3}$. 
The NBO analysis has been done using the CCSD(T) level of theory and aug-cc-pVTZ basis set \cite{ccpvtz} in the Gaussian09 software package.\cite{G09} MOs in figure \ref{homo-lumo}, has been generated after NBO analysis. A similar interaction between HOMO and LUMO did not observe in the rest of the systems.

A similar kind of partial overlapping in space (shown in figure \ref{homo-lumo}d) and interaction between HOMO and LUMO is observed in Li$^{-}$-H$_{2}$O. Here, the system stabilizes by 17.18$\times$2 = 34.36 kcal/mol by delocalizing Li-2s electron density to 2 O-H antibonding orbitals, drastically lowering the IP of HOMO to 0.42 eV.

In Li$^-$-NH$_3$ and Li$^-$-H$_2$O, we have seen delocalization of the electron density. However, despite the fact that both LiH-OH$^-$ and LiH-NH$_2^-$ are deprotonated systems, why is there no delocalization in these systems? This can be explained by the fact that lithium is less electronegative than oxygen and nitrogen in water and ammonia, respectively. As a result, the lone pair stays close to the oxygen/nitrogen atoms. Second, Li is bound to the p$_z$-orbitals of nitrogen and oxygen in LiH-OH$^-$ and LiH-NH$_2^-$. The electron density after deprotonation is still present in the p$_x$ and p$_y$ orbitals of nitrogen/oxygen. Unlike the s-orbital of Li-2s in Li$^-$-NH$_3$ and Li$^-$-H$_2$O, which is spherical, p-orbitals are directional. Therefore, in LiH-OH$^-$ and LiH-NH$_2^-$, we do not observe delocalization on deprotonation from ammonia/water.

Third, we noted that the IP was increased more by protonation from the metal center (IP's in LiH$_2^+$-NH$_3$) than by protonation from ammonia (IP's in LiH-NH$_4^+$). 
For example, the IP of HOMO is higher in LiH$_2^+$-NH$_3$ (18.24 eV) than in LiH-NH$_4^+$ (15.10 eV).  
The reason is the charge transfer. The location of a H$^+$ ion is critical within the molecule because it can help
promote and demote the charge transfer between lithium and nitrogen. Let's understand how the H$^+$ ion affects charge transfer. As discussed earlier, HOMO in LiH-NH$_3$ is more localized on 
hydrogen than lithium. An extra added H$^+$ ion shifts the lithium's 2s electron density. Thus, almost all-electron 
density will be on H$_2^+$ in LiH$_2^+$-NH$_3$ and LiH-NH$_4^+$, and lithium will be left with nearly nothing. 
However, the direct sharing of a lone pair of ammonia in LiH$_2^+$-NH$_3$ overcomes this lack of electron density on Li and stabilizes the system. LiH$_2^+$-NH$_3$ geometry shows that the additional H$^+$ ion is located far from the ammonia, facilitating charge transfer even more. Such direct electron sharing is impossible from ammonia to lithium in LiH-NH$_4^+$ since H$_2^+$  is between lithium and nitrogen. 
Fourth, we observe that the DIP value is the lowest (4.73 eV for Li$^-$-NH$_3$) and highest (48.26 eV for LiH$_2^+$-NH$_3$) when deprotonation and protonation occur from the metal center (LiH), respectively. That's because HOMO, which corresponds to the lowest DIP orbital, is a combination of lithium and hydrogen atomic orbitals. Thus, any change to these orbitals can lead to a considerable shift in DIP.
We have so far talked about how protonation and deprotonation affect IP and DIP values. We have observed that protonation increases the IP and DIP values, resulting in a few decay (inner valance) channels closing. For example, N-2s TBS in LiH$^+_2$-NH$_3$ and LiH-NH$_4^+$, and O-2s TBS in LiH$^+_2$-H$_2$O stop showing the decay. 
Note that the IP and DIP of the protonated/deprotonated system tell us whether a system is temporarily bound or not. In other words, if decay is possible or not. This means whether a lifetime of a decaying state will increase or decrease on protonation or deprotonation cannot be concluded based only on the IP and DIP values. To know that, we will now study how the protonation and deprotonation affect the decay width (in meV) and lifetime (in fs) of the TBS, along with the possible reasons.

\subsection*{Effect of protonation and deprotonation on the lifetime of Li-1s and O-2s/N-2s TBSs in LiH-H$_2$O and LiH-NH$_3$}
 
The visual representation of the ICD and ETMD processes for a general system ABC is shown in figure \ref{icd-etmd}. Since ICD and ETMD(2) include two atoms/molecules, we are not mentioning molecule C there. 
You can distinguish between various decay mechanisms by identifying their final DIP states. The final two-hole states are localized on two separate molecules in ICD, i.e., A$^+$B$^+$, but in ETMD(2), they are localized on one atom/molecules, i.e., AB$^{2+}$ or A$^{2+}$B. The two-hole localization in ETMD(3) is on the pair of neighboring molecules, i.e., C$^+$AB$^+$.
DIP eigenvalues and eigenvectors offer information on the localization of two holes in molecules. Specifically, eigenvalues tell us whether decay is feasible, whereas eigenvectors predict whether a process will be ICD, ETMD(2), or ETMD(3). 
In the following section, we will compare the systems in tables \ref{h2o} \& \ref{nh3} and examine potential explanations for any variations in the lifetime of Li-1s and O-2s/N-2s TBS on protonation and deprotonation. 

\begin{table*}
	\begin{center}
	\caption{Energetic position of the decaying state and the decay width after Li-1s and O-2s ionization of LiH-H$_{2}$O along with its protonated and deprotonated systems.}\label{h2o}
		\begin{tabular}{lcccc}	
\hline
\toprule		
System   &   \multicolumn{2}{c}{Li-1s}  &  \multicolumn{2}{c}{O-2s} \\
\cline{2-3} \cline{4-5}\\
& Energy(eV) & Width in meV (fs)  &  Energy (eV) & Width in meV (fs)   \\
\midrule
LiH-OH$^{-}$  &  55.56 & 64.27 (10.24) &  23.10 &   16.6 (39.62)    \\
LiH-H$_{2}$O  &   61.93 & 13.41 (49) &35.22  & 4.7 (139.6)     \\
LiH$_2^+$-H$_2$O &   70.79  & 50.17 (13.17) &   ---    &   ---   \\   
 \hline  
\bottomrule	
	\end{tabular}
	\end{center}
\end{table*}

\begin{figure}
  \includegraphics[width=\linewidth]{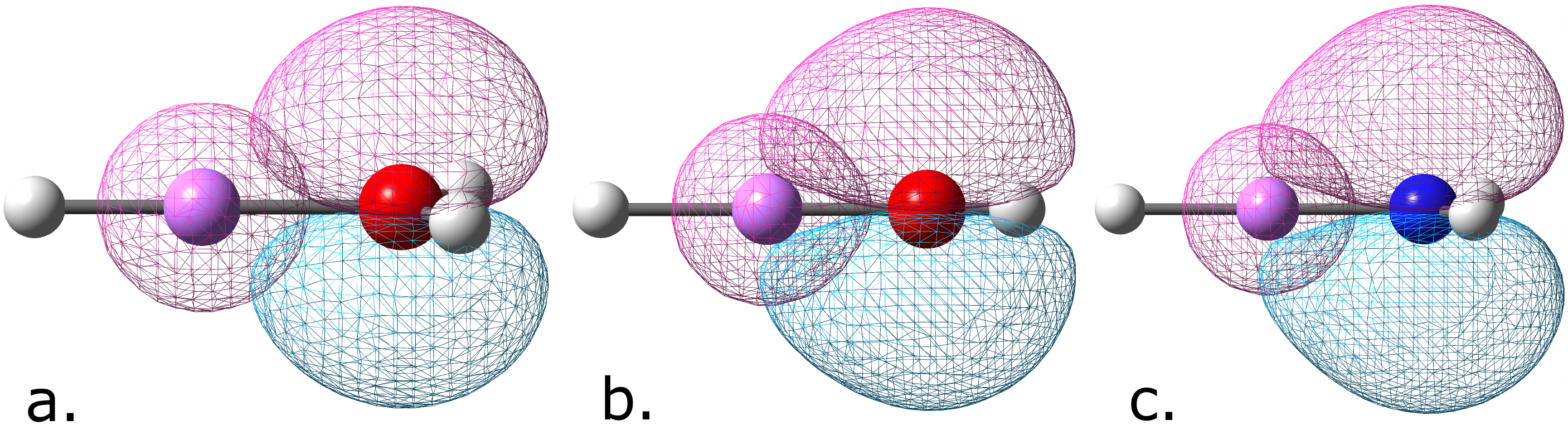}
  \caption{ a, b, and c show the partially overlapping in space between Li-1s orbital and the lone pair orbital of oxygen/nitrogen, which are perpendicular to the molecular axis in LiH-H$_2$O, LiH-OH$^-$, and LiH-NH$_2^-$, respectively. 
It shows that the orbital's partial overlapping in space increases from LiH-H$_2$O to LiH-OH$^-$.    
Images are generated for 0.02 e\AA$^{-3}$ isosurface value. Atom color code-: Red: Oxygen, Pink: Lithium, Blue: Nitrogen, and White: Hydrogen.}
  \label{overlapping}
\end{figure}

\subsubsection*{Lifetime of Li-1s TBS in LiH-H$_2$O, LiH-NH$_3$ and their deprotonated systems}

When we compare the lifetime of Li-1s TBS in deprotonated LiH-OH$^-$ and neutral LiH-H$_2$O systems in Table \ref{h2o}, we observe a decrease in the lifetime of Li-1s TBS. That's due to an increase in the orbital's partial overlapping in space (see fig. \ref{overlapping}a and \ref{overlapping}b) that facilitates a faster electron transfer between p-orbitals of oxygen to Li-1s orbital. 
This partial orbital overlap is influenced by three different causes. First is the charge. In LiH-OH$^-$, oxygen has an entirely negative charge, while lithium has a high positive charge. Here two adjacent atoms have opposite charges on them. We know opposite charges attract each other. Therefore, there will be an increase in attraction force (or partial overlapping of orbitals). 
The O-Li bond distance is the second. We observed a reduction in the O-Li bond length while switching from LiH-H$_2$O (1.924 {\AA}) to LiH-OH$^-$ (1.690 {\AA}). This further increases the partial overlapping between orbitals. 
The third is electron-electron repulsion. After LiH-H$_2$O deprotonated, an additional lone pair appeared on OH$^-$ in LiH-OH$^-$, which is perpendicular to the molecular plane. The Li-H bond's electron density is mostly on hydrogen (more electronegative than Li) because it is an ionic bond. 
Therefore, an incoming oxygen's electrons may feel substantially less repulsion from Li-2s electrons. This makes the shift of lone pair electrons of oxygen toward lithium more favorable. This shift of lone pairs increases further when there is a vacancy formation on Li during the decay, making electron transfer faster between oxygen and lithium. 
Thus, we are getting a low lifetime for Li-1s TBS in LiH-OH$^-$ than LiH-H$_2$O. Similar reasoning can explain a decrease in Li-1s TBS lifespan in LiH-NH$_2^-$ compared to LiH-NH$_3$.

 Additionally, we've seen that the decrease in Li-1s TBS lifetime in LiH-NH$_2^-$ relative to its neutral system is less pronounced than the decrease in Li-1s TBS lifetime in LiH-OH$^-$ relative to its neutral system. Two factors can explain this. First is the number of lone pairs. LiH-OH$^-$ has two lone pairs, which are perpendicular to the molecular plane, while LiH-NH$_2^-$ has one such lone pair. 
As a result, electron transfer from the oxygen/nitrogen lone pair to the Li-1s orbital will happen twice as quickly in LiH-OH$^-$ as in LiH-NH$_2^-$. Bond distance is the second. The O-Li bond distance (1.69 \AA) in LiH-OH$^-$ is shorter than the N-Li bond distance (1.85 \AA) in LiH-NH$_2^-$. 
Therefore, the partial overlapping between the lone pair orbital of oxygen and the lithium 1s orbital will be greater than that between the lone pair orbital of nitrogen and the lithium 1s orbital. This strong partial overlapping makes electron transfer faster between O/N and Li-1s in LiH-OH$^-$ than in LiH-NH$_2^-$. The lone pair of oxygen/nitrogen that we are talking about are the ones that are perpendicular to the molecular axis of each system. Hence, observing a significant reduction in LiH-OH$^-$ than in LiH-NH$_2^-$ compared to their neutral systems. 

\subsubsection*{Lifetime of O-2s and N-2s TBS in LiH-H$_2$O, LiH-NH$_3$ and their deprotonated systems}

We have seen a large shift in the lifetime of O-2s TBS and a minor decrease in N-2s TBS in  LiH-OH$^-$ and LiH-NH$_2^-$ compared to their respective neutral systems. There are three reasons for that. First, LiH-OH$^-$ has two lone pairs, which are perpendicular to the molecular plane, while LiH-NH$_2^-$ has one such lone pair. It means LiH-OH$^-$ and LiH-NH$_2^-$ each have one lone pair extra than their respective neutral systems. We are aware that bonded electron density is governed by two attractive forces from two separate nuclei, whereas lone pair electrons are affected by one nucleus (atom specific). It means the O-2s or N-2s vacancy can be filled faster in deprotonated systems than in respective neutral systems. Hence, we observed that the decay becomes faster, and the lifetime will reduce.
Second, oxygen is more electronegative than nitrogen. Thus, attraction felt by electrons is higher in LiH-OH$^-$ than LiH-NH$_2^-$, making decay faster. The third is the eigenvector analysis. We observe that N-2s have two possible decay channels in LiH-NH$_2^-$, while O-2s have four possible decay channels in LiH-OH$^-$. Based on eigenvector analysis, N-2s TBS decays by the ICD process only. In contrast, O-2s TBS decays by ETMD and ICD. 
Besides these three reasons, huge geometry and symmetry change may also be the reason for a large drop in the lifetime of O-2s in  LiH-OH$^-$ and a minor change in the lifetime of N-2s in LiH-NH$_2^-$ compared to their respective neutral systems. In water-related systems, the H-O-Li bond angle changes from 126.6$^{\circ}$ (in LiH-H$_2$O) to 180$^{\circ}$ (in LiH-OH$^-$), and symmetry changes from C$_{2V}$ to C$_{\infty v}$. While in ammonia-related systems, the H-N-Li bond angle changes from 112.4$^{\circ}$ (in LiH-NH$_3$) to 127.9$^{\circ}$ (in LiH-NH$_2^-$), and symmetry changes from C$_{3v}$ to C$_{2v}$. This change in the geometry of the ammonia-related system is not profound compared to the geometry change of LiH-OH$^-$ form LiH-H$_2$O.
 \begin{table*}
	\begin{center}
	\caption{Energetic position of the decaying state and the decay width after Li-1s and N-2s ionization of LiH-NH$_{3}$ along with its protonated and deprotonated systems.}\label{nh3}
		\begin{tabular}{lcccc}	
\hline
\toprule		
System   &   \multicolumn{2}{c}{Li-1s}  &  \multicolumn{2}{c}{N-2s} \\
\cline{2-3} \cline{4-5}\\
& Energy(eV) & Width in meV (fs)  &  Energy (eV) & Width in meV (fs)   \\
\midrule
LiH-NH$_3$ &   61.48  & 39.18  (16.79) & 30.00 & 7.19 (91.48)    \\
LiH$_2^+$-NH$_{3}$  & 70.17  &  10.39 (63.32) &    ---  &   ---     \\
 \hline
\multicolumn{5}{c}{protonation and deprotonation on nitrogen} \\
\hline
LiH-NH$_2^-$    &  55.50 &  65.52 (10.45) &   19.51 & 7.48 (87.90)   \\
LiH-NH$_3$ &   61.48  & 39.18 (16.79) &   30.00 & 7.19 (91.48)    \\
LiH-NH$_4^+$ & 71.01  & 144.06 (4.57) & ---   &   ---     \\
\hline
\bottomrule	
	\end{tabular}
	\end{center}
 \end{table*}

  \subsubsection*{Lifetime of Li-1s TBS in LiH-H$_2$O, LiH-NH$_3$ and their protonated systems}

The situation changes in LiH$^+_2$-H$_2$O and LiH$^+_2$-NH$_3$ because the molecule can be seen as three separate units, namely H$_2$O/NH$_3$, Li$^+$, and H$_2$. You might wonder if it makes a difference whether a molecule has two or three subunits. It matters because the prerequisites for the ETMD(3) process are met (two nearby atoms or molecules in addition to the atom or molecule that forms the TBS). 
The ETMD(3) decay process can be described for Li-1s TBS in LiH$^+_2$-H$_2$O and LiH$^+_2$-NH$_3$, as the lithium atom receives a second positive charge due to photoionization (Li-1s vacancy formation). Then an electron from the H$_2$ subunit will fill the Li-1s vacancy, and the released virtual photon will eject a secondary electron from either NH$_3$ or H$_2$O.
The directional nature of p-orbitals is critical in explaining increases and decreases in lifetime of the Li-1s TBS in the protonated systems (LiH$^+_2$-H$_2$O, LiH$^+_2$-NH$_3$, and LiH-NH$_4^+$).
The nitrogen/oxygen lone pair orbital is oriented toward the Li-1s TBS in LiH$^+_2$-NH$_3$ and perpendicular to the molecular plane in LiH$^+_2$-H$_2$O. As a result, charge transfer in LiH$^+_2$-NH$_3$ is eased, and we can observe that lone pairs of ammonia stabilize Li-1s TBS while LiH$^+_2$-H$_2$O does not experience this stabilization from water. This explains why the Li-1s TBS lifetime in LiH$^+_2$-H$_2$O and LiH$^+_2$-NH$_3$ differs from those of their neutral systems. 
Since H$_2$ sits between Li and ammonia in LiH-NH$_4^+$, there is no direct stabilizing effect by ammonia's lone pair in this compound. As a result, LiH-NH$_4^+$ has a lower Li-1s TBS lifetime than LiH-NH$_3$. 
Protonation raises the IP and DIP values, which shuts the decay channel for the O-2s and N-2s in their respective protonated systems (LiH$_2^+$-H$_2$O, LiH-NH$_4^+$, and LiH$_2^+$-NH$_3$). This explains the absence of N-2s and O-2s peaks in their respective protonated systems.
\section*{Conclusion}
\label{conclusion}
	
The impact of protonation and deprotonation on the system's IP, DIP, and lifetime has been examined in this paper.
LiH-NH$_3$ and LiH-H$_2$O have been selected as test systems. For our research, we employed the EOM-CCSD approach with CAP potential.
Our investigation found that protonation raises IP and DIP values relative to the neutral system, while deprotonation decreases IP and DIP values.
However, we do not see such a general trend for the decay width/lifetime on protonation or deprotonation. In general, a molecule's protonation or deprotonation impacts its structural stability, which results in noticeable changes to the system's geometry and charge distribution. 
For example, due to protonation/deprotonation, geometry changes dominate the decay widths in water-related systems. However, in ammonia-related systems, we observe that charge transfer makes the system stable (i.e., LiH$_2^+$-NH$_3$), causing a significant difference in decay width. 
These two elements influence the difference in the decay width of the system. We cannot state that protonation or deprotonation will cause an increase or decrease in decay width, unlike the IP/DIP values. 
We find that protonation (either from the metal center (LiH) or from H$_2$O/NH$_3$) shuts N-2s and O-2s (inner valence) decay pathways. However, the decay channel remains open for the core state Li-1s even after protonation. Li-1s TBS decays more quickly in deprotonated systems than in the corresponding neutral system because deprotonation increases the number of potential decay channels relative to the neutral system. Since the ammonia lone pair orbital is oriented toward Li and facilitates faster electron transmission, the Li-1s TBS decays in LiH-NH$_3$ more quickly than in LiH-H$_2$O. We draw the conclusion that the Li-1s TBS decays in neutral and deprotonated systems through ICD and ETMD(2) based on eigenvector analysis of the DIP. However, Li-1s decays via ETMD(2 \& 3) in LiH$_2^+$-NH$_3$ and LiH$_2^+$-H$_2$O following protonation.  O-2s and N-2s TBS decay via ICD and ETMD(2) in the neutral system. But in deprotonated systems, N-2s decay via ICD (in LiH-NH$_2^-$) only, whereas O-2s (in LiH-OH$^-$) decay via ICD and ETMD(2). All three decays are feasible for the decay of Li-1s in LiH-NH$_4^+$.
\section*{Theory and Computational Details}
\label{theory_comp}

We will initially examine the computational specifics before going through the decay processes' theory and procedures. 
Geometries of all molecules were optimized in the Gaussian09 software package \cite{G09}
using B3LYP \cite{b3lyp,B3,LYP,VWN} functional and 6-311++g(2d,p) basis set. \cite{O2H6311g} The IPs of
neutral, protonated, and deprotonated clusters are calculated using the equation of motion coupled-cluster
singles and doubles (EOM-CCSD) method. \cite{eomcc,eomcc1,eomcc2,eomcc3} The DIP values are computed using
the diagonalization of the 2-hole block of EOM-CCSD. The decay position and width of the system are calculated 
using the EOM-CCSD method augmented by the complex absorbing potential (CAP). \cite{cap1,cap2,cap3,cap4,cap5,cap6} 
We have used the aug-cc-pVTZ basis set \cite{ccpvtz} for CAP-EOM-CCSD calculations. For information on the CAP-EOM-CCSD technique for ICD, ETMD(2), and ETMD(3), 
see reference \cite{ravijctc}. 
For both neutral systems, the CAP-EOM-CCSD computations have been performed with various box sizes. The dimension of optimum box size was found to be 
C$_x$ = C$_y$ = C$_z$ = 9 a.u and ${\delta C}$ in protonated, deprotonated, and neutral LiH-NH$_3$. ${\delta C}$ = R/2 was added along the molecular axis.  
For protonated, deprotonated, and neutral LiH-H$_2$O systems, box dimensions were C$_x$ = C$_y$ = C$_z$ = 11 a.u and ${\delta C}$. 
Again, the ${\delta C}$ = R/2 value was added along the molecular axis. Calculations of IP, DIP, and decay widths are done using homegrown codes. We will now quickly go over the CAP-EOM-CCSD methodology. 


The target state, which is one hole state, is generated by the action of a linear operator $R(k)$ on the ground state wavefunction. The equation 
for the target state $|\Psi_{k}\rangle$ can be written as

\begin{eqnarray}
 |\Psi_{k}\rangle = R {(k)} |\Phi_{cc}\rangle
\end{eqnarray}

where $|\Phi_{cc} \rangle$ is the coupled-cluster wavefunction. It can be defined as 

\begin{eqnarray}
|\Phi_{cc}\rangle = e^{T} |\Phi_0\rangle
\end{eqnarray}

Where $|\Phi_0\rangle$ is the ground state Hartree-Fock wavefunction, $R(k)$ is an ionization operator, and T's (T = T$_1$ + T$_2$ for CCSD) are hole-particle excitation operators. T and R commute with each other. The linear operator R(k) within singles and doubles approximation is written as 

\begin{eqnarray}
R{(k)} = \sum_{i}r_{i}(k) i+ \sum_{i,j,a}r_{ij}^{a}(k) a^{\dagger}ji
\end{eqnarray}

The similarity transformed Hamiltonian $\overline{H}_{N}$ is formed as below.

\begin{eqnarray}
\overline{H}_{N}=e^{-T}H_{N}e^{T}- \langle \Phi_0| e^{-T}H_{N}e^{T} |\Phi_0 \rangle
\end{eqnarray}

The ionization potential values are obtained by diagonalizing the coupled-cluster similarity transformed Hamiltonian within $(N - 1)$ electron space. This similarity transformed Hamiltonian is formed within 1-hole and 2-hole one particle space. 

\begin{eqnarray}
\overline{H}_{N} R(k) |\Phi_0 \rangle = \omega_k R(k) |\Phi_0 \rangle
\end{eqnarray}

Where $\omega_k$ is the difference between the N and (N-1) electronic state, which is the IP of the k$^{th}$ state. 
The position and lifetime of the decaying state are obtained by augmenting the EOM-CCSD by complex absorbing potential (CAP).
CAP approach adds a one-particle potential to the physical Hamiltonian -i${\eta}$W. This makes the original Hamiltonian complex and non-hermitian H(${\eta}$) = H-i${\eta}$W. Where W is the real potential and ${\eta}$ is the strength of CAP. 
The complex eigenvalues of the non-hermitian Hamiltonian give us the system's position and half decay width. In CAP-EOM-CCSD, we get the total decay width, i.e., the contribution from all possible decay channels. We do not get the decay width for the individual decay process. 


 We solve the H(${\eta}$) = H-i${\eta}$W equation for various values of $\eta$ to obtain
complex energies that are $\eta$ dependent. Real part vs. imaginary part of energy plotted in energy plot that shows the $\eta$ trajectory. The stationary point on the trajectory provides the stabilization point.
%
%
\section*{Acknowledgment}

RK acknowledges financial support from Council of scientific and Industrial Research.
Authors acknowledge the computational facility at the CSIR-National Chemical Laboratory and IISER-Pune. 
Authors acknowledge Prof. L. S. Cederbaum for critical reading of the manuscript and the valuable suggestions.
\setlength{\bibsep}{0.0cm}
\bibliographystyle{wiley-chemistry}
\bibliography{icd_ph_b}



\section*{Entry for the Table of Contents}






\noindent\rule{11cm}{2pt}
\begin{minipage}{11cm}
\includegraphics[width=11cm]{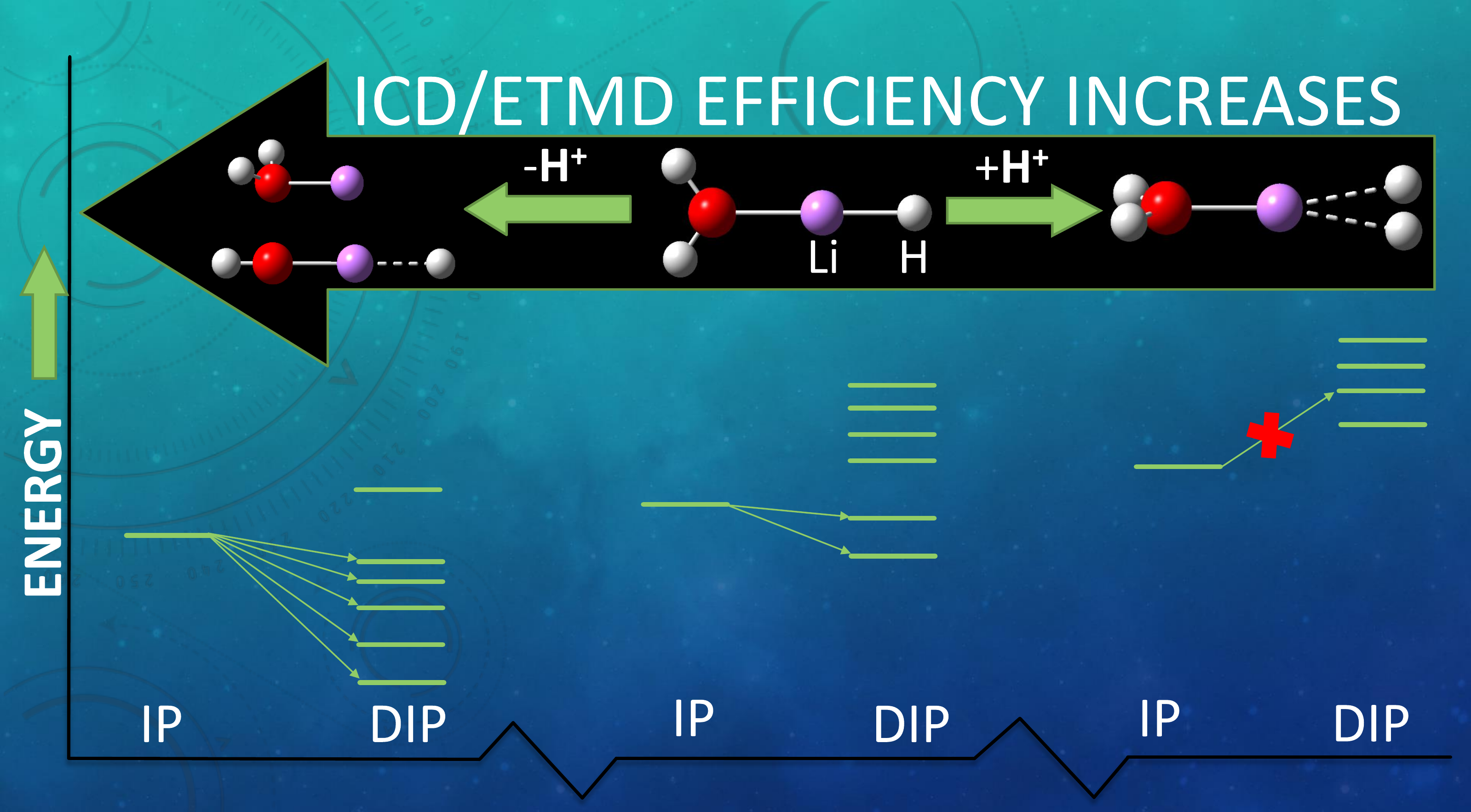}
\end{minipage}
\begin{minipage}{11cm}
\large\textsf{Authors should provide a short Table of Contents graphical abstract and accompanying text (up to 450 characters including spaces). The graphical abstract should stimulate curiosity. Repetition or paraphrasing of the title and experimental details should be avoided.}
\end{minipage}
\noindent\rule{11cm}{2pt}
\end{document}